\documentclass[conference,10pt]{IEEEtran}

\usepackage{graphics}
\usepackage{rotating}
\usepackage{amsfonts}
\usepackage{amsmath}
\usepackage{subfigure}
\usepackage{setspace}
\usepackage{amssymb}

\graphicspath{{Figures/}} \DeclareGraphicsExtensions{.eps,.ps}

\begin{document}

\title{Reduced Complexity Sphere Decoding for Square QAM via a New Lattice Representation} 

\author{\authorblockN{Luay Azzam and Ender Ayanoglu}\\
\authorblockA{
Department of Electrical Engineering and Computer Science\\
University of California, Irvine\\
Irvine, California 92697-2625\\
}}



\maketitle


\begin{abstract}

Sphere decoding (SD) is a low complexity maximum likelihood (ML)
detection algorithm, which has been adapted for different linear
channels in digital communications. The complexity of the SD has
been shown to be exponential in some cases, and polynomial in others
and under certain assumptions. The sphere radius and the number of
nodes visited throughout the tree traversal search are the decisive
factors for the complexity of the algorithm. The radius problem has
been addressed and treated widely in the literature. In this paper,
we propose a new structure for SD, which drastically reduces the
overall complexity. The complexity is measured in terms of the
floating point operations per second (FLOPS) and the number of nodes
visited throughout the algorithm's tree search. This reduction in
the complexity is due to the ability of decoding the real and
imaginary parts of each jointly detected symbol independently of
each other, making use of the new lattice representation. We further
show by simulations that the new approach achieves 80\% reduction in
the overall complexity compared to the conventional SD for a 2x2
system, and almost 50\% reduction for the 4x4 and 6x6 cases, thus
relaxing the requirements for hardware implementation.

\end{abstract}


%
%
\section{Introduction}

Minimizing the bit error rate (BER) and thus improving the
performance is the main challenge of receiver design for multiple-
input multiple-output (MIMO) systems. However, the performance
improvements usually come at the cost of increased complexity in the
receiver design. Assuming that the receiver has perfect knowledge of
the channel \textit{H}, different algorithms have been implemented
to separate the data streams corresponding to \textit{N} transmit
antennas \cite{Bolcskei}. Among these algorithms, Maximum Likelihood
detection (ML) is the optimum one. However, in MIMO systems, the ML
problem becomes exponential in the number of possible constellation
points making the algorithm unsuitable for practical purposes
\cite{Fettweis}. Sphere decoding, on the other hand, or the
Fincke-Pohst algorithm \cite{Fincke-Pohst}, reduces the
computational complexity for the class of computationally hard
combinatorial problems that arise in ML detection problems
\cite{Jalden-Ottersten}-\cite{Hassibi-Vikalo}.

Complexity reduction techniques for SD have been proposed in the
literature. Among these techniques, the increased radius search
(IRS) \cite{Viterbo-Boutros} and the improved increasing radius
search (IIRS) \cite{Zhao-Giannakis} suggested improving SD
complexity efficiency by making a good choice of the sphere radius,
trying to reduce the number of candidates in the search space. The
former suggested a set of sphere radii ${c_{1}<c_{2}<...<c_{n}}$
such that SD starts with ${c_{1}}$ trying to find a candidate. If no
candidates were found, SD executes again using the increased radius
${c_{2}}$. The algorithm continues until either a candidate is found
or the radius is increased to ${c_{n}}$ which should be large enough
to guarantee obtaining at least one candidate. Whereas, the latter
provided a mechanism to avoid the waste of computations taking place
in the former method when a certain radius ${c_{m}}$ does not lead
to a candidate solution. Obviously, these two techniques studied the
complexity problem from the radius choice perspective.

In this paper we improve the SD complexity efficiency by reducing
the number of FLOPS required by the SD algorithm keeping in mind the
importance of choosing a radius. The radius should not be too small
to result in an empty sphere and thus restarting the search, and at
the same time, it should not be too large to increase the number of
lattice points to be searched. We use the formula presented in
\cite{Rekaya-Belfiore} for the radius, which is ${\textit{d}^{2} =
2\sigma^{2}\textit{N}}$, where \textit{N} is the problem dimension
and ${\sigma^{2}}$ is the noise variance. The reduction of the
number of FLOPS is accomplished by introducing a new and proper
lattice representation, as well as incorporating quantization at a
certain level of the SD search. It is also important to mention that
searching the lattice points using this new formulation can be
performed in parallel, since the new proposed structure in this
paper enables decoding the real and imaginary parts of each symbol
independently and at the same time.

The remainder of this paper is organized as follows: In Section
\ref{sec:problem_definition}, a problem definition is introduced and
a brief review of the conventional SD algorithm is presented. In
Section \ref{sec:New_Rep}, we propose the new lattice representation
and perform the mathematical derivations for complexity reduction.
Performance and complexity comparisons for different number of
antennas or modulation schemes are included in Section
\ref{sec:results}. Finally, we conclude the paper in Section
\ref{sec:conclusion}.

\section{Problem Definition and the Conventional Sphere Decoder}\label{sec:problem_definition}

Consider a {MIMO} system with {\textit{N}} transmit and {\textit{M}}
receive antennas. The received signal at each instant of time is
given by
\begin{equation}
y = Hs + v
 \label{eq:received_y}
\end{equation}
where $y \in \mathbb{C}^{\textit{M}}$, $H \in
\mathbb{C}^{\textit{M}{\mathrm{x}}\textit{N}}$ is the channel
matrix, $s \in \mathbb{C}^{\textit{N}}$ is an \textit{N} dimensional
transmitted complex vector whose entries have real and imaginary
parts that are integers, $v \in \mathbb{C}^{\textit{M}}$ is the
i.i.d complex additive white Gaussian noise {(AWGN)} vector with
zero-mean and covariance matrix ${\sigma^{2}{I}}$. Usually, the
elements of the vector $s$ are constrained to a finite set $\Omega$
where $\Omega\subset\mathbb{Z}^{2N}$, e.g., ${\Omega = \{-3, -1, 1,
3\}}^{2N}$ for 16-QAM (quadrature amplitude modulation) where
$\mathbb{Z}$ and $\mathbb{C}$ denote the sets of integers and
complex numbers respectively.

Assuming $H$ is known at the receiver, the ML detection is given by
\begin{equation}
\hat{s} = \textrm{arg} \min\limits_{s \in \Omega} {||y - Hs||}^{2}.
 \label{eq:received_yhat}
\end{equation}
Solving (\ref{eq:received_yhat}) becomes impractical and exhaustive
for high transmission rates, and the complexity grows exponentially.
Therefore, instead of searching the whole space defined by all
combinations drawn by the set $\Omega$, SD solves this problem by
searching only over those lattice points or combinations that lie
inside a sphere centered around the received vector $y$ and of
radius $d$. Introducing this constraint on (\ref{eq:received_yhat})
will change the problem to
\begin{equation}
\hat{s} = \textrm{arg} \min\limits_{s \in \Omega} {||y - Hs||}^{2} <
d^{2}. \label{eq:SD_constraint}
\end{equation}
A frequently used solution for the QAM-modulated complex signal
model given in (\ref{eq:SD_constraint}) is to decompose the
$N$-dimensional problem into a 2$N$-dimensional real-valued problem,
which then can be written as
\begin{equation}
\left[\begin{array} {c c} \Re\{y\}\\ \Im\{y\}
\end{array}
\right] = \left[\begin{array} {c c} \Re\{H\} & -\Im\{H\} \\
\Im\{H\} & \Re\{H\} \end{array} \right] \left[\begin{array} {c c}
\Re\{s\}\\ \Im\{s\} \end{array} \right] + \left[\begin{array} {c c}
\Re\{v\}\\ \Im\{v\} \end{array} \right]
 \label{eq:real_dec}
\end{equation}
where $\Re\{y\}$ and $\Im\{y\}$ denote the real and imaginary parts
of $y$, respectively \cite{Bolcskei}, \cite{Jalden-Ottersten},
\cite{Rekaya-Belfiore}-\cite{Chan-Lee}. Assuming $N=M$ in the
sequel, and introducing the QR decomposition of $H$, where
\textit{R} is an upper triangular matrix, and the matrix \textit{Q}
is unitary, (\ref{eq:SD_constraint}) can be written as
\begin{equation}
\hat{s} = \textrm{arg} \min\limits_{s \in \Omega} {||\bar{y} -
Rs||}^{2} < d^{2} \label{eq:received_ydash}
\end{equation}
where $\bar{y}=Q^{H}y$. Let
\textit{R}=$\left[r_{i,j}\right]_{2\textit{N}\mathrm{x}2\textit{N}}$
and note that \textit{R} is upper triangular. Now to solve
(\ref{eq:received_ydash}), the SD algorithm constructs a tree, where
the branches coming out of each node correspond to the elements
drawn by the set $\Omega$. It then executes the decoding process
starting with the last layer ($l=2N$) which matches the first level
in the tree, calculating the partial metric
$||\Im\{\bar{y}_{N}\}-r_{2N,2N}\Im\{s_{N}\}||^{2}$, and working its
way up in a similar way to the successive interference cancelation
technique, until decoding the first layer by calculating the
corresponding partial metric
$||\Re\{\bar{y}_{1}\}-r_{1,1}\Re\{s_{1}\}+ ... +
\Re\{\bar{y}_{N}\}-r_{1,N}\Re\{s_{N}\}+
\Im\{\bar{y}_{1}\}-r_{1,N+1}\Im\{s_{1}\}+...+
\Im\{\bar{y}_{N}\}-r_{1,2N}\Im\{s_{N}\}||^{2}$. The summation of all
partial metrics along the path of a node starting from the root
constitutes the weight of that node. If that weight exceeds the
square of the sphere radius $\textit{d}^{2}$, the algorithm prunes
the corresponding branch, declaring it as an improbable way to a
candidate solution. In other words, all nodes that lead to a
solution that is outside the sphere are pruned at some level of the
tree. Whenever a valid lattice point at the bottom level of the tree
is found within the sphere, the square of the sphere radius
$\textit{d}^{2}$ is set to the newly found point weight, thus
reducing the search space for finding other candidate solutions.
Finally, the leaf with the lowest weight will be the survivor one,
and the path along the tree from the root to that leaf represents
the estimated solution $\hat{s}$.

To this end, it is important to emphasize the fact that the
complexity of this algorithm, although it is much lower than the ML
detection, is still exponential at low SNR, and is directly related
to the choice of the radius $d$, as well as the number of floating
point operations taking place at every tree node inside the sphere.


\section{New Lattice Representation}\label{sec:New_Rep}

The lattice representation given in (\ref{eq:real_dec}) imposes a
major restriction on the tree search algorithm. Specifically, the
search has to be executed serially from one level to another on the
tree. This can be made clearer by writing the partial metric weight
formula as
\begin{equation}
w_{l}(x^{(l)}) = w_{l+1}(x^{(l+1)}) +
|\hat{y}_{l}-\sum_{k=l}^{2N}r_{l,k}x_{k}|^2
 \label{eq:partial_weight}
\end{equation}
with $l=2N,2N-1,\ldots,1$, $w_{2N+1}(x^{(2N+1)}{)=0}$ and where
$\{x_{1},x_{2},...,x_{N}\}$, $\{x_{N+1},x_{N+2},...,x_{2N}\}$ are
the real and imaginary parts of $\{s_{1},s_{2},...,s_{N}\}$
respectively.

Obviously, the SD algorithm starts from the upper level in the tree
($l=2N$), traversing down one level at a time, and computing the
weight for one or more nodes (depending on the search strategy
adopted, i.e., depth-first, breadth-first, or other reported
techniques in the literature) until finding a candidate solution at
the bottom level of the tree ($l=1$). According to this
representation, it is impossible, for instance, to calculate
$\sum_{k=l}^{2N}r_{l,k}x_{k}$ for a node in level ($l=2N-1$) without
assigning an estimate for $x_{2\textit{N}}$. This approach results
in two related drawbacks. First, the decoding of any
$x_{\textit{l}}$ requires an estimate value for all preceding
$x_{j}$ for $j=l+1,...,2N$. Secondly, there is no room for parallel
computations since the structure of the tree search is sequential.

The main contribution in this paper is that we relax the tree search
structure making it more flexible for parallelism, and at the same
time reducing the number of computations required at each node by
making the decoding of every two adjacent levels in the tree totally
independent of each other.

We start by reshaping the channel matrix representation given in
(\ref{eq:real_dec}) in the following form:
\begin{equation} \tilde{H}{=}
\left[\begin{array} {c c c c c} \Re(H_{1,1}) & -\Im(H_{1,1}) &
\cdots & \Re(H_{1,N}) & -\Im(H_{1,N}) \\ \Im(H_{1,1}) & \Re(H_{1,1})
&  \cdots &  \Im(H_{1,N}) & \Re(H_{1,N}) \\ \vdots & \vdots & \ddots
& \vdots & \vdots \\ \Re(H_{N,1}) & -\Im(H_{N,1}) & \cdots &
\Re(H_{N,N}) & -\Im(H_{N,N}) \\ \Im(H_{N,1}) & \Re(H_{N,1}) & \cdots
&  \Im(H_{N,N}) & \Re(H_{N,N}) \end{array} \right]
 \label{eq:newH}
\end{equation}
where $H_{m,n}$ is the i.i.d. complex path gain from transmit
antenna $n$ to receive antenna $m$. By looking attentively at the
columns of $\tilde{H}$ starting from the left hand side, and
defining each pair of columns as one set, we observe that the
columns in each set are orthogonal, a property that has a
substantial effect on the structure of the problem. Using this
channel representation changes the order of detection of the
received symbols to the following form
\begin{equation} \hat{y}=
\left[\begin{array} {c c c c c} \Re(\hat{y}_{1}) & \Im(\hat{y}_{1})
& \cdots & \Re(\hat{y}_{N}) & \Im(\hat{y}_{N})\end{array}
\right]^{T}.
 \label{eq:new_yhat}
\end{equation}
This means that the first and second levels of the search tree
correspond to the real and imaginary parts of $s_{N}$, unlike the
conventional SD, where those levels correspond to the imaginary part
of $s_{N}$ and $s_{N-1}$ respectively. The new structure becomes
advantageous after applying the QR decomposition to $\tilde{H}$. By
doing so, and due to that special form of orthogonality among the
columns of each set, all the elements $r_{k,k+1}$ for
{$k=1,3,...,2N-1$} in the upper triangular matrix \textit{R} become
zero. The locations of these zeros are very important since they
introduce orthogonality between the real and imaginary parts of
every detected symbol.

In the following, we will prove that the QR decomposition of
$\tilde{H}$ introduces the aforementioned zeros. There are several
methods for computing the QR decomposition, we will do so by means
of the Gram-Schmidt algorithm.

\noindent \begin{proof} Let
\begin{equation} \tilde{H}=
\left[\begin{array} {c c c c}
\mathbf{\tilde{\textbf{h}_{\textit{1}}}} &
\mathbf{\tilde{\textbf{h}_{\textit{2}}}} & \cdots &
\mathbf{\tilde{\textbf{h}_{\textit{2N}}}}
\end{array} \right] \label{eq:columns_H}
\end{equation}
where $\mathbf{\tilde{\textbf{h}_{\textit{k}}}}$ is the $k$th column
of $\tilde{\textit{H}}$. Recalling the Gram-Schmidt algorithm, we
define

\noindent $\mathbf{\textbf{u}}_\textit{1} =
\mathbf{\tilde{\textbf{h}}_{\textit{1}}}$ \\
\noindent and then, \vspace{3pt} \\
\noindent $\mathbf{\textbf{u}}_\textit{k} =
\mathbf{\tilde{\textbf{h}}_{\textit{k}}}-\sum_{j=1}^{k-1}\mathrm{\phi}_{\mathbf{\textbf{u}}_\textit{j}}\,\mathbf{\tilde{\textbf{h}}_{\textit{k}}}$
\indent\indent\indent\indent\indent\indent\indent\ for
$k=2,3,...,2N$. \\ 
\noindent where
$\mathrm{\phi}_{\mathbf{u}_\textit{j}}\,\mathbf{\tilde{h}_{\textit{k}}}$
is the projection of vector $\mathbf{\tilde{h}_{\textit{k}}}$ onto
${\mathbf{u}_\textit{j}}$ defined by
\begin{align}
{\mathbf{\phi}_{\mathbf{\textbf{u}}_\textit{j}}\,\mathbf{\tilde{\textbf{h}}_{\textit{k}}}}
=
{\langle\mathbf{\tilde{\textbf{h}}}_\textit{k},\mathbf{\tilde{\textbf{u}}_{\textit{j}}}\rangle\over\langle\mathbf{\tilde{\textbf{u}}_{\textit{j}}},\mathbf{\tilde{\textbf{u}}_{\textit{j}}}\rangle}\mathbf{\tilde{\textbf{u}}_{\textit{j}}}
\end{align}
\noindent and $\mathbf{\textbf{e}}_\textit{k} =
{\mathbf{\textbf{u}}_\textit{k}\over\|\mathbf{\textbf{u}}_\textit{k}\|}$
for $k=1,2,\ldots,2N$. Rearranging the equations \\
$\mathbf{\tilde{h}_{\textit{1}}} =
\mathbf{e}_\textit{1}\|\mathbf{u}_\textit{1}\|$\\
$\mathbf{\tilde{h}_{\textit{2}}} =
\mathrm{\phi}_{\mathbf{u}_\textit{1}}\,\mathbf{\tilde{h}_{\textit{2}}}+\mathbf{e}_\textit{2}\|\mathbf{u}_\textit{2}\|$\\
$\mathbf{\tilde{h}_{3}} =
\mathrm{\phi}_{\mathbf{u}_\textit{1}}\,\mathbf{\tilde{h}_{\textit{3}}}+\mathrm{\phi}_{\mathbf{e}_\textit{2}}\,\mathbf{\tilde{h}_{\textit{3}}}+\mathbf{e}_\textit{3}\|\mathbf{u}_\textit{3}\|$\\
$\vdots$ \\
$\mathbf{\tilde{h}_{\textit{k}}} =
\sum_{j=1}^{k-1}\mathrm{\phi}_{\mathbf{u}_\textit{j}}\,\mathbf{\tilde{h}_{\textit{k}}}+\mathbf{e}_\textit{k}\|\mathbf{u}_\textit{k}\|$. \\
Now, writing these equations in the matrix
form, we get: 
\begin{align}
\left[\begin{array} {c c c} \mathbf{e}_\textit{1} & \cdots &
\mathbf{e}_\textit{n}
\end{array} \right]
\left[\begin{array} {c c c c c}
\|\mathbf{u}_\textit{1}\| & \langle\mathbf{e}_\textit{1},\mathbf{\tilde{h}_{\textit{2}}}\rangle &  \langle\mathbf{e}_\textit{1},\mathbf{\tilde{h}_{\textit{3}}}\rangle  & \ldots \\
0                & \|\mathbf{u}_\textit{2}\|                        &  \langle\mathbf{e}_\textit{2},\mathbf{\tilde{h}_{\textit{3}}}\rangle  & \ldots \\
0                & 0                                       & \|\mathbf{u}_\textit{3}\|                          & \ldots \\
\vdots           & \vdots                                  & \vdots
& \ddots \end{array} \right]
\end{align}
Obviously, the matrix to the left is the orthogonal unitary
\textit{Q} matrix, and the one to the right is the upper triangular
\textit{R} matrix. \\
\noindent Now our task is to show that the terms $
\langle\mathbf{e}_\textit{k},\mathbf{\tilde{h}_{\textit{k+1}}}\rangle$
are zero for $k=1,3,\ldots,2N-1$. Three observations
conclude the proof. \\ 
\noindent First, since $\mathbf{\tilde{h}_{\textit{k}}}$ and
$\mathbf{\tilde{h}_{\textit{k+1}}}$ are orthogonal for
$k=1,3,\ldots,2N-1$, then
$\mathrm{\phi}_{\mathbf{\textbf{u}}_\textit{k}}\,\mathbf{\tilde{\textbf{h}}_{\textit{k+1}}}$
=
$\mathrm{\phi}_{\mathbf{\textbf{u}}_\textit{k+1}}\,\mathbf{\tilde{\textbf{h}}_{\textit{k}}}$
= 0 for the same $k$. \vspace{2pt} \\
\noindent Second, the projection of $\mathbf{\textbf{u}}_\textit{m}$
for $m=1,3,\ldots,k-2$ on the columns
$\mathbf{\tilde{h}_{\textit{k}}}$ and
$\mathbf{\tilde{h}_{\textit{k+1}}}$ respectively is equal to the
projection of $\mathbf{\textbf{u}}_\textit{m+1}$ on the columns
$\mathbf{\tilde{h}_{\textit{k+1}}}$ and
-$\mathbf{\tilde{h}_{\textit{k}}}$ respectively. To formalize this:
\begin{align}\nonumber
\langle\mathbf{u_{\textit{m}}},\mathbf{\tilde{h}_{\textit{k}}}\rangle
=\langle\mathbf{u_{\textit{m+1}}},\mathbf{\tilde{h}_{\textit{k+1}}}\rangle
\nonumber \\
\langle\mathbf{u_{\textit{m}}},\mathbf{\tilde{h}_{\textit{k+1}}}\rangle
= -
\langle\mathbf{u_{\textit{m+1}}},\mathbf{\tilde{h}_{\textit{k}}}\rangle
\end{align}
\noindent for $k=1,3,\ldots,2N-1$ and $m=1,3,\ldots,k-2$. This
property becomes obvious by using the first observation and
revisiting the special structure of
(\ref{eq:newH}).\vspace{2pt} \\
\noindent Third, making use of the first two observations, and
noting that
$||\mathbf{\tilde{\textbf{h}}_{\textit{k}}}||$=$||\mathbf{\tilde{\textbf{h}}_{\textit{k+1}}}||$
for $k=1,3,\ldots,2N-1$, it can be easily shown that
$||\mathbf{\textbf{u}_{\textit{k}}}||$=$||\mathbf{\textbf{u}_{\textit{k+1}}}||$
for the same $k$. \\
\noindent Then,
\begin{align}
\langle\mathbf{e}_\textit{k},\mathbf{\tilde{h}_{\textit{k+1}}}\rangle
 = &
 {\langle{\mathbf{u}_\textit{k}\over\|\mathbf{u}_\textit{k}\|},\mathbf{\tilde{h}_{\textit{k+1}}}\rangle} \nonumber \\
 = &
 {1\over\|\mathbf{u}_k\|}\langle{\mathbf{\tilde{\textbf{h}}_{\textit{k}}}-\sum_{j=1}^{k-1}\mathrm{\phi}_{\mathbf{\textbf{u}}_\textit{j}}\,\mathbf{\tilde{\textbf{h}}_{\textit{k}}},\mathbf{\tilde{h}_{\textit{k+1}}}\rangle} \nonumber \\
 = & {1\over\|\mathbf{u}_k\|}({\langle{\mathbf{\tilde{\textbf{h}}_{\textit{k}}}}, \mathbf{\tilde{\textbf{h}}_{\textit{k+1}}}\rangle}-  {{\langle{\mathbf{\tilde{\textbf{h}}_{\textit{k}}}},\mathbf{u}_\textit{1} \rangle}{\langle{\mathbf{u}_\textit{1}},\mathbf{\tilde{\textbf{h}}_{\textit{k+1}}} \rangle}\over\langle{\mathbf{u}_\textit{1},\mathbf{u}_\textit{1}} \rangle}- \nonumber\\ & {{\langle{\mathbf{\tilde{\textbf{h}}_{\textit{k}}}}, \mathbf{u}_\textit{2}\rangle}{\langle{\mathbf{u}_\textit{2}},\mathbf{\tilde{\textbf{h}}_{\textit{k+1}}} \rangle}\over\langle{\mathbf{u}_\textit{2}},\mathbf{u}_\textit{2} \rangle}- \cdots- \nonumber\\ & {{\langle{\mathbf{\tilde{\textbf{h}}_{\textit{k}}}}, \mathbf{u}_\textit{k-2}\rangle}{\langle{\mathbf{u}_\textit{k-2}},\mathbf{\tilde{\textbf{h}}_{\textit{k+1}}} \rangle}\over\langle{\mathbf{u}_\textit{k-1}},\mathbf{u}_\textit{k-1} \rangle}- {{\langle{\mathbf{\tilde{\textbf{h}}_{\textit{k}}}}, \mathbf{u}_\textit{k-1}\rangle}{\langle{\mathbf{u}_\textit{k-1}},\mathbf{\tilde{\textbf{h}}_{\textit{k+1}}} \rangle}\over\langle{\mathbf{u}_\textit{k-2}},\mathbf{u}_\textit{k-2} \rangle})
\label{eq:inner_product}
\end{align}
\noindent Now, applying the above observations to
(\ref{eq:inner_product}), we get
\begin{align}
\langle\mathbf{e}_\textit{k},\mathbf{\tilde{h}_{\textit{k+1}}}\rangle
 = &
{1\over\|\mathbf{u}_k\|}(0 -
{{\langle{\mathbf{\tilde{\textbf{h}}_{\textit{k}}}},\mathbf{u}_\textit{1}
\rangle}{\langle{\mathbf{u}_\textit{1}},\mathbf{\tilde{\textbf{h}}_{\textit{k+1}}}
\rangle}\over\|\mathbf{u}_\textit{1}\|^2}- \nonumber\\ &
{-{\langle{\mathbf{u}_\textit{1}},\mathbf{\tilde{\textbf{h}}_{\textit{k+1}}}
\rangle}{{\langle{\mathbf{\tilde{\textbf{h}}_{\textit{k}}}},\mathbf{u}_\textit{1}
\rangle}}\over\|\mathbf{u}_\textit{1}\|^2}- \cdots- \nonumber\\ &
{{\langle{\mathbf{\tilde{\textbf{h}}_{\textit{k}}}},
\mathbf{u}_\textit{k-2}\rangle}{\langle{\mathbf{u}_\textit{k-2}},\mathbf{\tilde{\textbf{h}}_{\textit{k+1}}}
\rangle}\over\|\mathbf{u}_\textit{k-1}\|^2}- \nonumber
{-{{\langle{\mathbf{u}_\textit{k-2}},\mathbf{\tilde{\textbf{h}}_{\textit{k+1}}}
\rangle}}{\langle{\mathbf{\tilde{\textbf{h}}_{\textit{k}}}},
\mathbf{u}_\textit{k-2}\rangle}\over\|\mathbf{u}_\textit{k-1}\|^2})
\nonumber\\ & = 0 \nonumber \label{eq:inner_product}
\end{align}
This concludes the proof.
\end{proof}
In this context, the SD algorithm executes in the following way.
First, the partial metric weight
$|\hat{y}_{2N}-r_{2N,2N}\hat{x}_{2N}|^2$ for the $\mu$ nodes in the
first level of the tree is computed, where $\mu$ is the number of
elements in $\Omega$. This metric is then checked against the
specified sphere radius $d^2$. If the weight at any node is greater
than the sphere radius then the corresponding branch is pruned.
Otherwise, the metric value is saved for the next step. At the same
time, another set of $\mu$ partial metric computations of the form
$|\hat{y}_{2N-1}-r_{2N-1,2N-1}\hat{x}_{2N-1}|^2$ take place at the
second level, since these two levels are independent as proved
above. These metrics are checked against $d^2$ in a similar way to
that done in the above level. The weights of the survivor nodes from
both levels are summed up and the summation is checked against the
sphere constraint, ending up with a set of survivor ${\hat{s}_{N}}$
symbols. Secondly, the estimation of the remaining $\hat{x}_{2N-2}$
or $\hat{s}_{N-1}$ symbols is done by quantization to the nearest
constellation element in $\Omega$. In other words, the values of
$\hat{x}_{2N-2},\ldots,\hat{x}_{1}$ are calculated recursively for
each combination of survived $\hat{x}_{2N},\hat{x}_{2N-1}$, and the
total weight given by $||\hat{y}-R\hat{s}||^2$ is determined at the
bottom level of the tree for those leaves that obey the radius
constraint. Finally, the leaf with the minimum weight is chosen to
be the
decoded message $(\hat{s})$. This can be formalized as\\
\noindent \underline{Step1:} \\
\noindent for $i=1$ to $\mu$ \\
\indent $\hat{x}_{2N}=\Omega(\mu)$ \\
\indent $\hat{x}_{2N-1}=\Omega(\mu)$ \\
\indent  if $|\hat{y}_{2N}-r_{2N,2N}\hat{x}_{2N}|^2 < d^2$ $\rightarrow$ add to survivor set 1 \\
\indent  else $\rightarrow$ prune branch \\
\indent  if $|\hat{y}_{2N-1}-r_{2N-1,2N-1}\hat{x}_{2N-1}|^2 < d^2$ $\rightarrow$ add to survivor\\
\indent set 2 \\
\indent  else $\rightarrow$ prune branch \\
\noindent next $i$ \\
\noindent save all combinations of $\hat{x}_{2N},\hat{x}_{2N-1}$
whose weight summations comply to the radius constraint. Denote the
number of survivors at the end of this step by \{$\lambda$\}. \\
\noindent \underline{Step2:} (for every combination in $\lambda$, calculate $\hat{x}_{2N-2},...,\hat{x}_{1}$ recursively as shown below) \\
\noindent for $l=2N-2$ to $1$, step $-2$ \\
\indent set $\upsilon$=$l/2$, and calculate \\
\begin{align}  \nonumber
e_{l} = \sum_{\stackrel{k=l+1}{}}^{2N}r_{l,k}\hat{x}_{k}, & \qquad
e_{l-1} =
\sum_{\stackrel{k=l+1}{}}^{2N}r_{l-1,k}\hat{x}_{k}\\\nonumber\\
\hat{x}_{l}=\mho({\Im(\hat{y}_{\upsilon})-e_{l}\over r_{l,l}}), &
\qquad \hat{x}_{l-1}=\mho({\Re(\hat{y}_{\upsilon})-e_{l-1}\over
r_{l-1,l-1}}) \nonumber
\end{align}
\begin{align} \nonumber
w_{l}(\hat{x}^{(l)}) & = w_{l+1}(\hat{x}^{(l+1)}) +
|\Im{(\hat{y}_{\upsilon})}-e_{l}-\hat{x}_{l}r_{l,l}|^2 \nonumber \\
w_{l-1}(\hat{x}^{(l-1)}) & = w_{l}(\hat{x}^{(l)}) +
|\Re{(\hat{y}_{\upsilon})}-e_{l-1}-\hat{x}_{l-1}r_{l-1,l-1}|^2
\nonumber
\end{align}
\noindent next $l$ \\
\noindent where $\mho$(.) quantizes the value (.) to the closest
element in the set $\Omega$. The output of the above two steps is a
set of candidate solutions $\hat{x}_{2N},...,\hat{x}_{1}$ with
corresponding weights. \\
\noindent \underline{Step3:} \\
\noindent choose that set of $\hat{x}_{2N},...,\hat{x}_{1}$ which
has the lowest weight to be the detected message.

Finally, the above algorithm's complexity is linear with the number
of antennas, and the performance is optimal for MIMO systems having
two antennas at both ends. However, this performance becomes
suboptimal for systems with $N\geq3$ (e.g., there is a $4$ dB loss
compared to the conventional SD at a BER of $10^{-5}$ for a 4x4
system ). This is mainly due to the use of quantization which takes
place at all tree levels except the first two, and makes the
estimation of $\hat{x}$s loose as we further traverse down in the
tree. Thus, we introduce minor heuristic rules in the middle levels
of the tree when $N\geq3$, while still using the above steps at the
very first and very last two levels in the tree, in order to obtain
near optimal performance (less than $1$ dB loss), sticking with a
complexity that is very much small compared to the conventional SD.
A brief discussion on how to specify these rules are proposed in
Section \ref{sec:results}.



\section{Simulation Results}\label{sec:results}

We have considered 2x2, 4x4, and 6x6 cases using 16-QAM and 64-QAM
modulation schemes. As mentioned in the previous section, we
introduce heuristic rules in the middle levels of the tree when
$N\geq3$. Therefore, in our simulations for the 4x4 and 6x6 cases,
we executed the algorithm in the following way:

For the 4x4 system, the first two levels of the tree which
correspond to the imaginary and real parts of the symbol
$\hat{s}_{4}$ are treated the same way as explained in Step 1 of the
algorithm. For each survivor $\hat{s}_{4}$, the weight for all
different $\mu^2$ possibilities of $\hat{s}_{3}$  is calculated, and
those weights that violate the radius constraint are dismissed
(${d^{2} = 2\sigma^{2}N}$ \cite{Rekaya-Belfiore}). The best 8
survivors, or in other words, those 8 $\hat{s}_{3}$'s that have
lowest weights are kept for next steps while for the others the
corresponding paths are pruned. In the third two levels, the same
procedure performed in the previous step is applied and the best 8
$\hat{s}_{2}$'s are kept. Finally, a quantization process followed
by an estimation of the transmitted message is carried out exactly
the same way as in step 2 and 3 explained in the previous section.
On the other hand, the 6x6 case has similar approach but with
different parameters. The first two levels are treated similarly as
explained in step 1. For the 16-QAM (64-QAM) case, the best 16, 8,
and 4 (32,32, and 16) survivors of $\hat{s}_{6}$, $\hat{s}_{5}$, and
$\hat{s}_{4}$ respectively are kept in the middle levels until
reaching the last four levels which are then processed by
quantization, in order to obtain $\hat{s}_{2}$ and $\hat{s}_{1}$.

\begin{figure}[h]
{\includegraphics[width=1\linewidth]{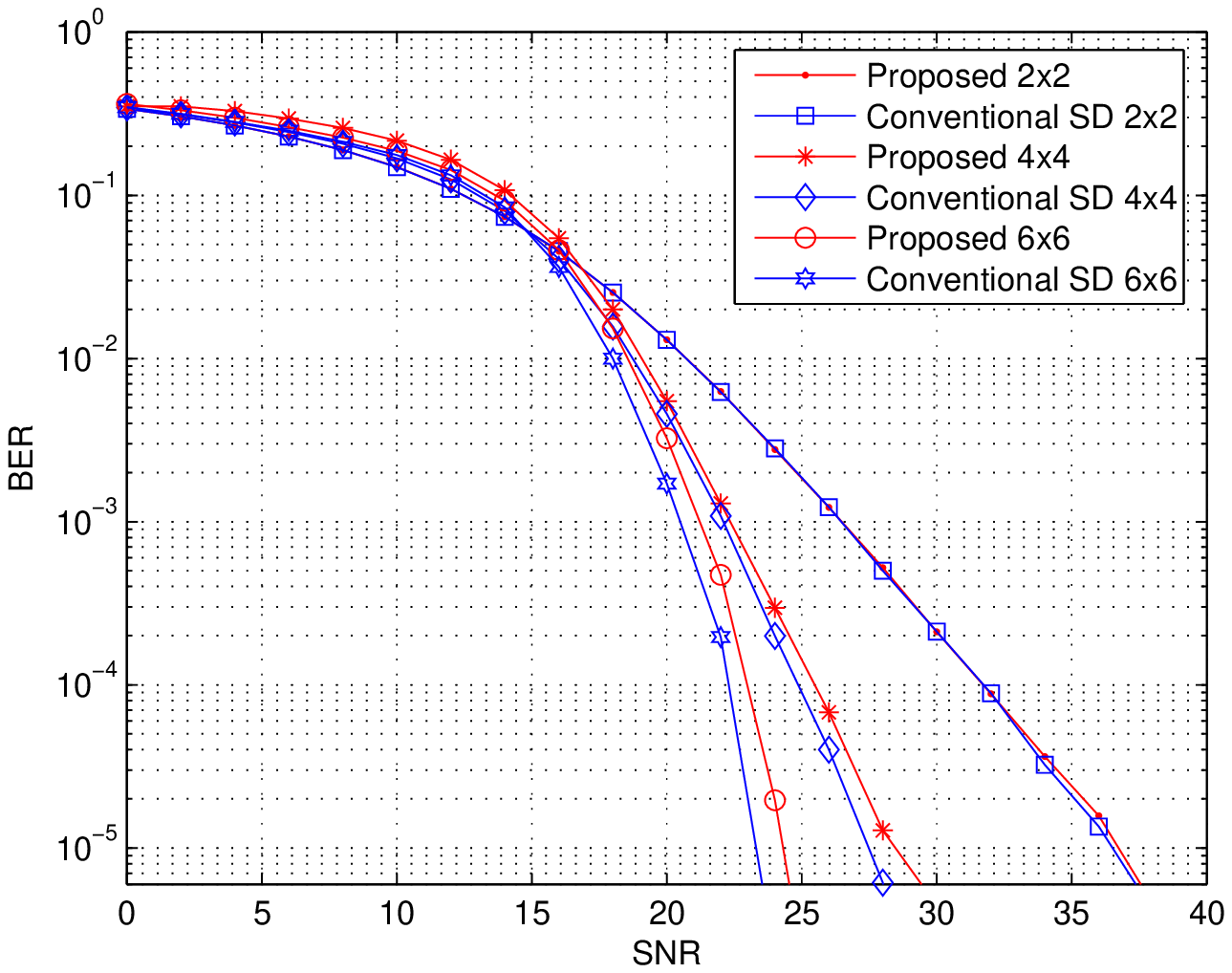}}
\caption{BER vs SNR for the proposed and conventional SD over a 2x2, 4x4, and 6x6 MIMO flat fading channel using 16-QAM modulation.}\ 
\label{fig:16qam_Perf} {\includegraphics[width=1\linewidth]{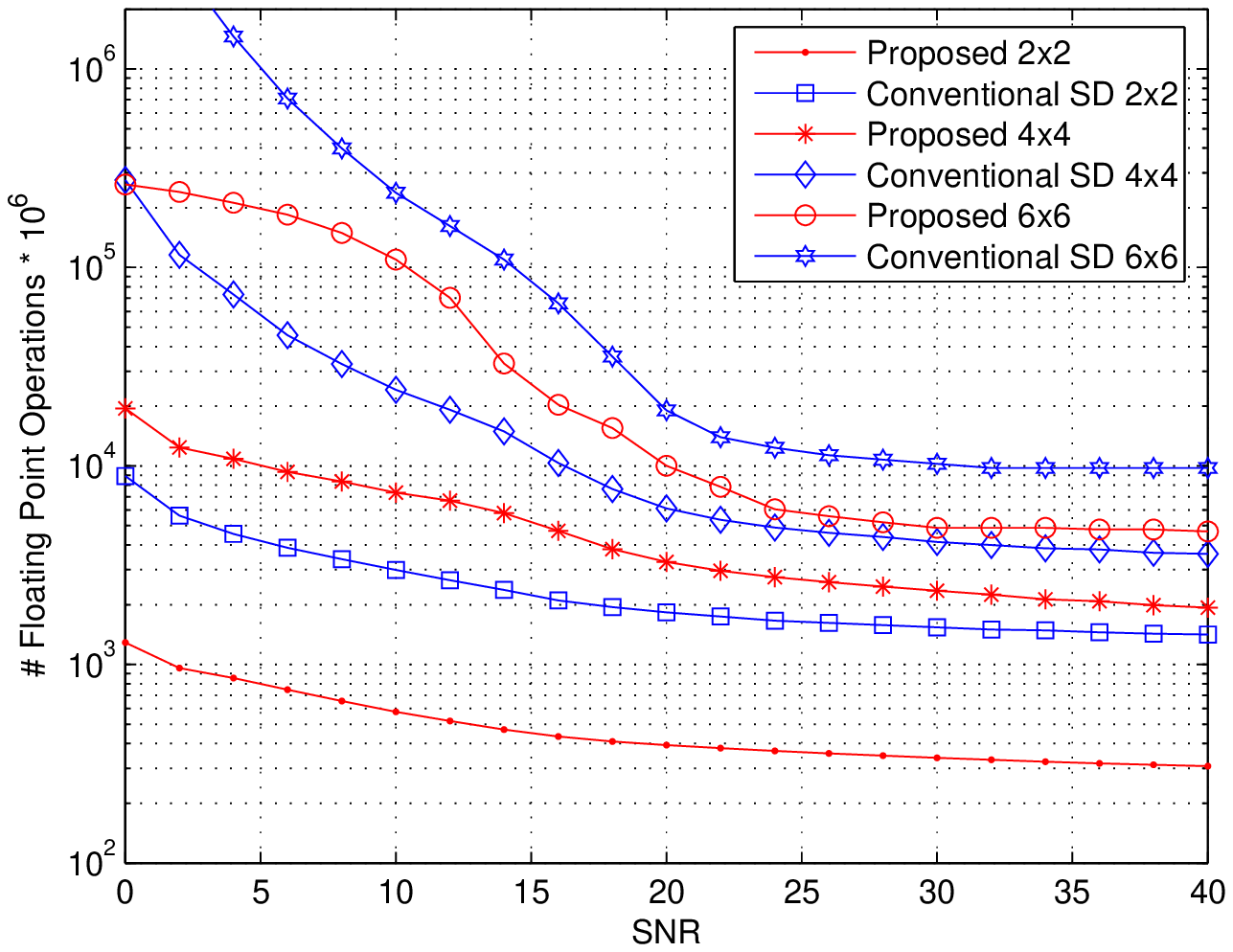}} \caption{Total number of floating
point operations vs SNR for the proposed and conventional SD over a 2x2, 4x4 and 6x6 MIMO flat fading channel using
16-QAM modulation.}\label{fig:16qam_FLOPS}
\end{figure}

\begin{figure}[t]
\centering 
{\includegraphics[width=1\linewidth]{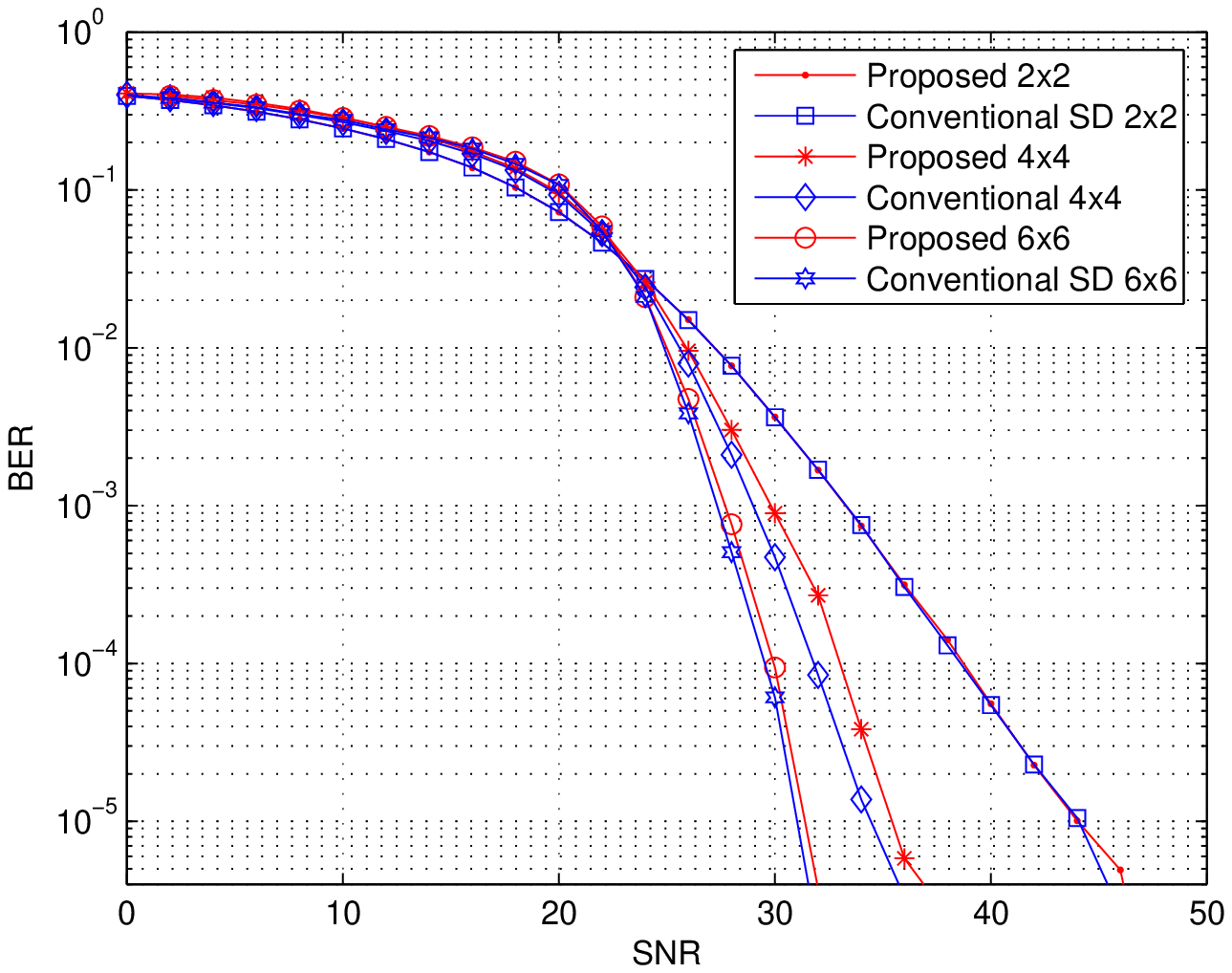}}
\caption{BER vs SNR for the proposed and conventional SD over a 2x2, 4x4, and 6x6 MIMO flat fading channel using 64-QAM modulation.}\ 
\label{fig:64qam_Perf} {\includegraphics[width=1\linewidth]{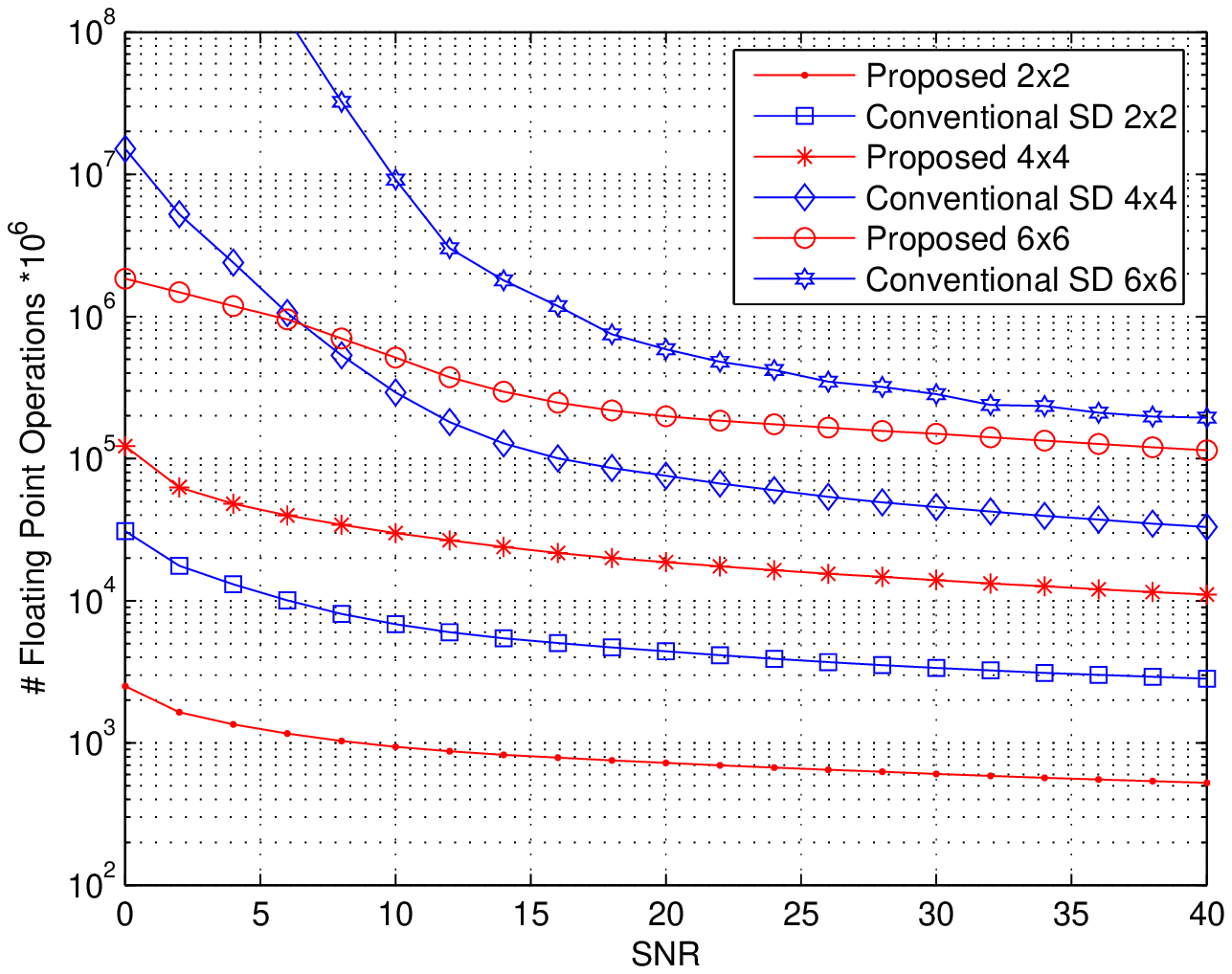}} \caption{Total number of floating
point operations vs SNR for the proposed and conventional SD over a 2x2, 4x4, and 6x6 MIMO flat fading channel using
64-QAM modulation.}\label{fig:64qam_FLOPS}
\end{figure}

Figure \ref{fig:16qam_Perf} reports the performance of the proposed
algorithm versus the conventional SD, for 2x2, 4x4, and 6x6 cases
using 16-QAM modulation. We observe that the proposed algorithm
achieves exactly the same performance as the conventional SD, but
with much smaller complexity as shown in Figure
\ref{fig:16qam_FLOPS}. However, there is almost $0.5-1$ dB
performance loss in the proposed 4x4 and 6x6 compared to the
conventional. This loss is due to the $k$-best criteria adoption at
a certain level of the tree as well as applying the quantization
process at the low levels of the tree as mentioned above. From
Figure \ref{fig:16qam_FLOPS}, it is clear that the proposed
algorithm reduces the complexity by 80\% for the 2x2 case, and 50\%
for both the 4x4 and 6x6 systems.
Figures \ref{fig:64qam_Perf} and \ref{fig:64qam_FLOPS} show the
performance and complexity curves for the 2x2, 4x4, and 6x6 cases,
for the 64-QAM modulation. Again, the performance is shown to be
close to the conventional for the 2x2 case, and has almost $0.5-1$
dB degradation loss for the 4x4 and 6x6 cases. The difference in the
complexity for the proposed and conventional SD are within the same
range as in the 16-QAM modulation case.


\section{Conclusions}\label{sec:conclusion}

A simple and general lattice representation in the context of sphere
decoding was proposed in this paper. The performance of the proposed
structure was shown to be optimal for 2x2 systems while close to
optimal ($0.5-1$ dB loss) in the 4x4 and 6x6 cases. A complexity
reduction of 80\% was attainable for the 2x2 case, and 50\% for the
4x4 and 6x6 cases, compared to their correspondence for the
conventional SD.



\bibliographystyle{IEEEtran}
\bibliography{IEEEabrv,C:/luay/Bibliography/VLSI,C:/luay/Bibliography/SD,C:/enis/bibliography/books,C:/enis/bibliography/MyBib,C:/enis/bibliography/space-time-codes,C:/enis/bibliography/bicm,C:/enis/bibliography/STC-OFDM}

\end{document}